\documentclass[sn-nature,Numbered]{sn-jnl}
\usepackage{graphicx}%
\usepackage{multirow}%
\usepackage{amsmath,amssymb,amsfonts}
\usepackage{amsthm}%
\usepackage{mathrsfs}%
\usepackage[title]{appendix}%
\usepackage{textcomp}%
\usepackage{manyfoot}%
\usepackage{booktabs}%
\usepackage{algorithm}%
\usepackage{algorithmicx}%
\usepackage{algpseudocode}%
\usepackage{listings}%
\usepackage[left]{lineno}
\usepackage{newtxtext}  
\usepackage{kotex}
\usepackage[dvipsnames]{xcolor}
\usepackage[capitalise,nameinlink]{cleveref}
\usepackage[super]{natbib}
\usepackage{hyperref}
\usepackage{color,soul}
\theoremstyle{thmstyleone}%
\theoremstyle{thmstyletwo}%
\theoremstyle{thmstylethree}%
\begin{document}

\title{\normalfont\fontsize{14}{16}\selectfont\textbf{Underwater Willis lens for broadband low-frequency focusing}}

\author[1]{\fnm{Beomseok} \sur{Oh}}
\author[1]{\fnm{Dongwoo} \sur{Lee}}
\author[2]{\fnm{Yeon-Seong} \sur{Choo}}
\author[2]{\fnm{Sung-Hoon} \sur{Byun}}
\author[3]{\fnm{Jehyeon} \sur{Shin}}
\author*[2]{\fnm{Sea-Moon} \sur{Kim}}
\email{smkim@kriso.re.kr}
\author*[1,3,4,5,6]{\fnm{Junsuk} \sur{Rho}}
\email{jsrho@postech.ac.kr}

\affil[1]{\orgdiv{Department of Mechanical Engineering}, \orgname{Pohang University of Science and Technology (POSTECH)}, \orgaddress{\city{Pohang}, \postcode{37673}, \country{Republic of Korea}}}

\affil[2]{\orgdiv{Ocean and Maritime Digital Technology Research Division}, \orgname{Korea Research Institute of Ships $\&$ Ocean Engineering (KRISO)}, \orgaddress{\city{Daejeon}, \postcode{34103}, \country{Republic of Korea}}}

\affil[3]{\orgdiv{Graduate School of Artificial Intelligence}, \orgname{Pohang University of Science and Technology (POSTECH)}, \orgaddress{\city{Pohang}, \postcode{37673}, \country{Republic of Korea}}}

\affil[4]{\orgdiv{Department of Chemical Engineering}, \orgname{Pohang University of Science and Technology (POSTECH)}, \orgaddress{\city{Pohang}, \postcode{37673}, \country{Republic of Korea}}}

\affil[5]{\orgdiv{Department of Electrical Engineering}, \orgname{Pohang University of Science and Technology (POSTECH)}, \orgaddress{\city{Pohang}, \postcode{37673}, \country{Republic of Korea}}}

\affil[6]{\orgname{POSCO-POSTECH-RIST Convergence Research Center for Flat Optics and Metaphotonics}, \orgaddress{\city{Pohang}, \postcode{37673}, \country{Republic of Korea}}}

\nolinenumbers
\abstract{Broadband underwater sound focusing in the low-frequency range is essential for various applications such as battery-free environmental monitoring and sensing. However, achieving low-frequency underwater focusing typically necessitates bulky, heavy structures that hinder practical deployment. 
Here, we introduce a three-dimensional underwater lens comprising cavity-based locally resonant asymmetric structures, enabling the efficient manipulation of low-frequency waterborne sound through a densely packed lattice configuration. 
We experimentally validated its broadband focusing performance over a range of 20--35 kHz. In addition, we observed that our lens exhibits asymmetric backscattering\textemdash a distinctive effect arising from its bianisotropic nature\textemdash which we term the Willis lens. Unlike conventional underwater lenses that rely on fully filled structures, our design employs cavity-based scatterers, achieving a lighter yet robust focusing performance. With its lightweight, efficient, and reliable design, the Willis lens provides a promising platform for underwater sensor networks and future advancements in on-demand waterborne sound focusing.
}

\maketitle
\clearpage
\section*{Introduction}
The precise manipulation of underwater acoustic waves, particularly within the low-frequency range that includes the audible spectrum, is essential for a wide range of applications, such as sound navigation and ranging (SONAR), acquisition of oceanographic data, underwater communication, monitoring of marine pollution, and tactical surveillance systems \cite{akyildiz2005underwater,felemban2015underwater,elfes1987sonar,andre2011listening}. 
Acoustic waves are more practical underwater owing to the significant attenuation of electromagnetic waves in aquatic environments \cite{hovem2007underwater,sherman2007transducers}. 
From this perspective, controlling underwater acoustic energy is critical, for instance, in power-free sensor networks wherein batteries have a limited capacity and are often not rechargeable \cite{afzal2022battery}. However, this task is more challenging underwater than in air because underwater devices are vulnerable to harsh environmental conditions, such as fouling and corrosion. In this context, passive structure-based underwater sound focusing can play a crucial role in a wide range of applications, such as long-term deep-sea monitoring with wireless acoustic energy transfer, even under such extreme conditions. Recent advances in acoustic wave manipulation technologies using metamaterials and phononic crystals have enabled precise control of waves with high degrees of freedom \cite{cummer2016controlling,assouar2018acoustic,oh2023engineering,lee2022piezoelectric,vasileiadis2021progress,lee2024wide,dong2023underwater}. Despite extensive research on adapting metamaterials for underwater sound manipulation \cite{dong2023underwater,zhou2024underwater,li2024janus, qu2022underwater,lee2021underwater,yang2023experimental}, experimental demonstrations of three-dimensional lenses for low-frequency underwater focusing remain scarce. Notably, a significant milestone in this field was the first experimental realization of an acoustic gradient-index (GRIN) lens, achieved as recently as 2014 \cite{zhao2023review} and enabled by advanced additive manufacturing techniques. Although Luneburg’s seminal work in electromagnetics dates back to 1944 \cite{luneburg1966mathematical}, its application in acoustics has only recently been realized. 
Over the past decade, various experimental implementations have emerged that have driven further advancements in the field of acoustics. Furthermore, existing studies remain limited despite the significant potential in the low-frequency range and the necessity for aquatic applications, such as battery-free monitoring, ecological acoustic recording, and on-demand wireless power transfer. 
Most efforts have predominantly focused on two-dimensional lenses (or three-dimensional lenses in high-frequency regimes) \cite{zhao2023review}, primarily because they require bulky structures, and experimental validation is more challenging underwater than in air. Additionally, the reliance on solid scatterer-based structures \cite{allam20203d,kim2022three,xie2018acoustic,li2021focus,lu2021grin,kim2021acoustic,kim2021poroelastic,tong20233d}, which utilize Bragg scattering, contributes to the bulkiness of these systems, posing challenges for their practical implementation.

In this paper, we propose a three-dimensional underwater lens designed with bianisotropic scatterers, which refer to as the Willis lens. Our lens demonstrates a robust focusing performance across the 20--35 kHz range; this was achieved by employing locally resonant asymmetric structures with high spatial resolution, as opposed to conventional solid scatterer-based lenses. This approach enables a cost-effective and lightweight design, offering significant potential for practical underwater sound manipulation. Through comprehensive experimental validation and numerical simulations, we confirm the effectiveness of our Willis lens design in achieving broadband and wide-angle focusing behavior. Moreover, the bianisotropic (pressure--velocity cross-coupling) properties \cite{quan2018maximum,sieck2017origins,esfahlani2021homogenization,muhlestein2017experimental,melnikov2019acoustic,su2018retrieval,sepehrirahnama2022willis,li2018systematic,muhlestein2016reciprocity,demir2024equivalence,wen2023acoustic} inherent in spatially asymmetric unit scatterers lead to asymmetric reflection characteristics. Consequently, we observed angle-invariant broadband focusing along with incidence-dependent backscattering. To the best of our knowledge, this study entailed the first experimental realization of three-dimensional and wide-angle underwater focusing using locally resonant asymmetric building blocks, particularly in the low-frequency domain. These results show that our approach can address the primary challenges pertaining to underwater lenses that are crucial for practical applications, namely, broadband operation, lens size and weight constraints, and robust signal-to-noise ratio (SNR) in submerged environments.

\section*{Results}
\subsubsection*{Design principle and characterization of the underwater Willis lens}
\begin{figure}[!h]
\centering
\includegraphics [width=0.99\textwidth] {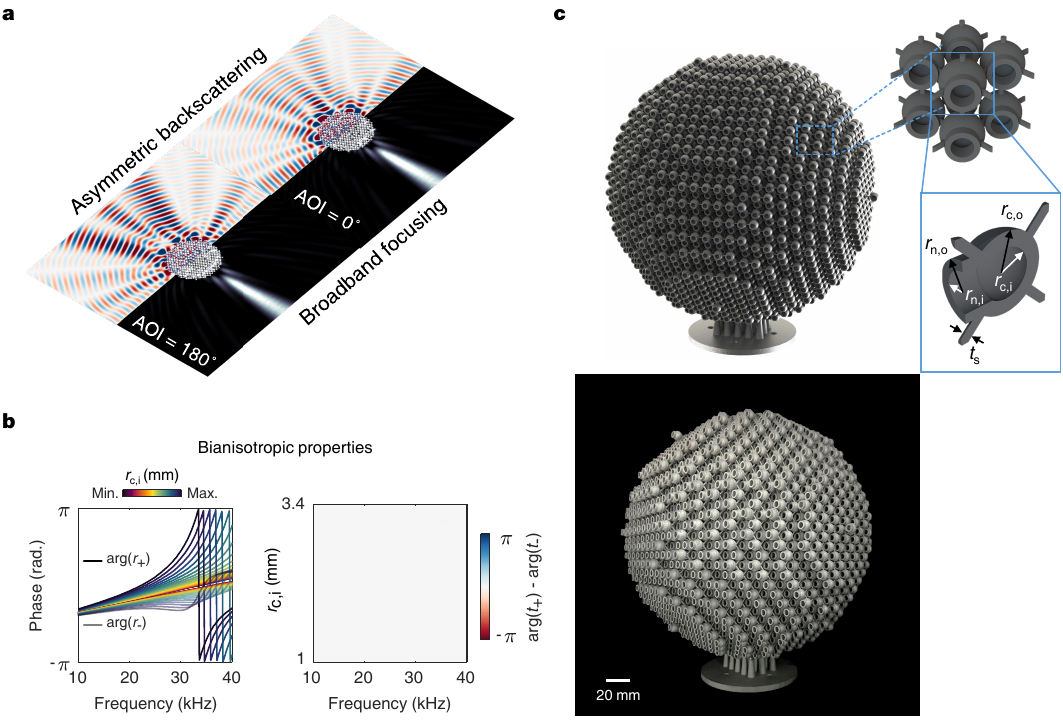}
\caption{\textbf{Concept and principle of the proposed underwater Willis lens.} 
\textbf{a} Schematic of the proposed Willis lens, featuring broadband underwater sound focusing and asymmetric backscattering attributed to pressure--velocity cross-coupling (i.e., acoustic Willis coupling).
\textbf{b} Bianisotropic properties of Willis scatterers: 
Left, asymmetric reflection phase as a function of the geometric parameter (shown in (c)) and frequency.
Right, phase differences in transmission for different incident directions with respect to the geometric parameter and frequency. The ``$+$'' and ``$-$'' symbols indicate incident angles of 0° and 180°, respectively.
\textbf{c} Top: Rendered images of the proposed underwater lens. Each unit cell features an asymmetric geometry resembling a Helmholtz resonator. The geometric parameters include the radii of the outer and inner necks ($r_\textnormal{n,o}$, $r_\textnormal{n,i}$), the radii of the outer and inner cavities ($r_\textnormal{c,o}$, $r_\textnormal{c,i}$), and the thickness of connecting supporters ($t_\textnormal{s}$).
Bottom: A photograph of the fabricated Willis lens (scale bar: 20 mm; total diameter: 240 mm).}
\label{fig1}
\end{figure}
A conceptual representation of this study is shown in \cref{fig1}a. The main features of our lens are its broadband focusing in the low-frequency regime, which is independent of the incidence direction, and its asymmetric backscattering, which is incidence-dependent. To realize these characteristics, we designed a lens based on asymmetric scatterers shaped similarly to Helmholtz resonators (\cref{fig1}c). By introducing spatially asymmetric structures instead of traditional symmetric scatterers (e.g., sphere, cube, and cross shapes), we induce the pressure--velocity cross-coupling known as bianisotropy or Willis coupling \cite{quan2018maximum,sieck2017origins,esfahlani2021homogenization,muhlestein2017experimental,melnikov2019acoustic,su2018retrieval,sepehrirahnama2022willis,li2018systematic,muhlestein2016reciprocity,demir2024equivalence,wen2023acoustic}. 
In particular, Helmholtz resonator-like asymmetric building blocks typically exhibit propagation  direction-independent transmission properties while featuring incidence-dependent reflection characteristics \cite{esfahlani2021homogenization}. 
In this configuration, the acoustic cavity acts as a secondary scatterer that can be used to adjust the overall scattering response \cite{sepehrirahnama2022willis}. 
Numerous previous studies have been conducted on electromagnetic, acoustic, and elastic bianisotropic metamaterials based on these characteristics \cite{quan2018maximum,sieck2017origins,esfahlani2021homogenization,muhlestein2017experimental,melnikov2019acoustic,su2018retrieval,sepehrirahnama2022willis,li2018systematic,muhlestein2016reciprocity,demir2024equivalence,wen2023acoustic,asadchy2018bianisotropic,kriegler2009bianisotropic}.
Notably, our unit scatterers exhibited asymmetric reflection properties depending on the incident direction (left panel of \cref{fig1}b) while maintaining a configuration-independent consistent transmission characteristic, as indicated by the gray-colored region in the right panel of \cref{fig1}b, where the phase difference remains close to zero. By incorporating these features into the lens design, we achieve consistent focusing performance regardless of the incidence direction, while observing distinct asymmetric backscattering induced by Willis coupling, which varies with the configuration direction. \Cref{fig1}c depicts the designed and fabricated Willis lens for underwater sound focusing, along with the geometric details of the unit scatterers. Because water, the background medium, has a high acoustic characteristic impedance, we used stainless steel (STS 316L) as the base material. We then fabricated the lens as a monolithic, one-body structure using a stereolithography-based metal additive manufacturing process (see Methods and Supplementary Note 1 for details). The lens has a diameter of 240 mm and an operating frequency range of approximately 20--35 kHz. 

Owing to the nature of the underwater environment, maximizing the contrast in the characteristic impedance between water and the unit cells is more challenging than in air. While bubble- or trapped-air-inclusion-based unit cells can achieve a large impedance contrast (and thus a high refractive index), they are inherently more sensitive and vulnerable to external environmental factors such as hydrostatic pressure. 
From a practical standpoint, particularly when considering durability under harsh submerged conditions (e.g., high hydrostatic pressure and exposure to drag forces when mounted on ships or similar platforms), it is more suitable to construct unit cells using high-density materials such as steel. Considering these factors, designing a high-index unit scatterer (similar to those used in optical and acoustic metamaterials for airborne applications) remains challenging in underwater environments. Consequently, a gradient-index (GRIN) lens, which can provide broadband focusing with a relatively low-index distribution, is well-suited for underwater applications.
Existing GRIN lenses have primarily been designed based on the Bragg scattering properties of phononic crystals; this implies that the period (or lattice constant) of each unit cell is inevitably close to the wavelength scale \cite{allam20203d,kim2022three,xie2018acoustic,li2021focus,lu2021grin,kim2021acoustic,kim2021poroelastic,tong20233d}. 
In real-world low-frequency applications with constrained lens sizes, relatively large unit cells pose challenges for achieving an ideal continuous index profile. 
This limitation results in low spatial resolution and pronounced discreteness, leading to reduced focusing efficiency and parasitic diffraction \cite{pan2022dielectric}. 
This issue is related to impedance matching, wherein an increased number of layers reduces the intensity loss by enabling a more gradual change in the effective parameters.
\begin{figure}[!t]
\centering
\includegraphics [width=0.99\textwidth] {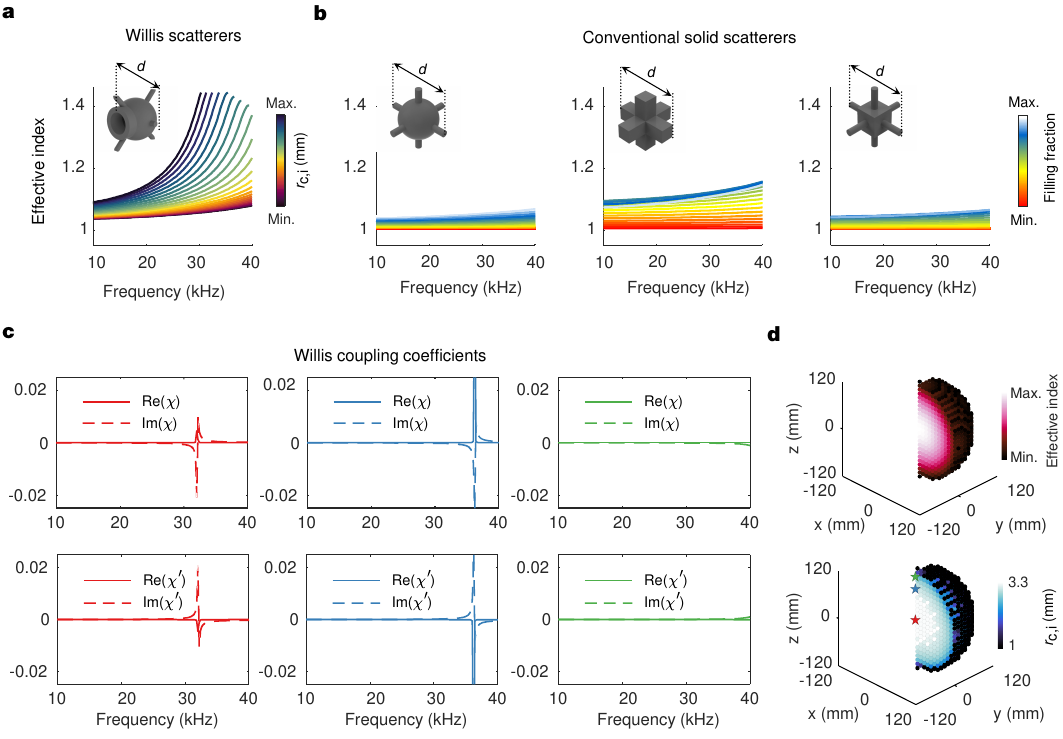}
\caption{\textbf{Physical characterization of Willis scatterers.}
\textbf{a, b} Effective refractive indices for (a) the proposed Willis scatterers and (b) conventional solid scatterers, including sphere (left), plus (middle) and cube (right) shapes. The parameter \textit{d} represents the period of the unit structure, which is identical for all configurations.
\textbf{c} Willis coupling coefficients of representative asymmetric Willis scatterers, where geometrical values are indicated by star symbols in (d).
\textbf{d} Effective index distribution of the designed underwater Willis lens (top) and corresponding geometrical parameters (bottom).
}
\label{fig2}
\end{figure}
Beyond the simple asymmetric property discussed previously, our unit scatterers exhibit locally resonant characteristics rather than relying on conventional fully filled scatterers. By leveraging their deep subwavelength nature, as is widely recognized in locally resonant metamaterials \cite{liu2000locally,ma2016acoustic}, we can design the lens to achieve the desired effective index in the low-frequency range while maintaining a high spatial resolution. \Cref{fig2}a and b illustrate the effective index profiles of the proposed unit structure, compared to conventional scatterers: sphere, plus, and cube shapes, all designed with the same periodicity. The period \textit{d} = 0.01 m, corresponding to approximately $0.18\,\lambda$ at the center frequency; this implies that each unit structure is on the deep subwavelength scale. 
Unlike conventional solid scatterers, Willis scatterers can achieve the desired refractive index in the low-frequency range despite having a relatively small lattice constant. This, in turn, leads to an increased focusing efficiency within the limited lens size, thereby enhancing the SNR in practical environments where background noise is inherently more prominent.

As aforementioned, our unit scatterers exhibit bianisotropic properties due to their spatial asymmetry. For a general Willis-coupled unit cell with a subwavelength lattice constant, the scattering properties can be understood as the coupling of an acoustic monopole and three orthogonal dipole moments \cite{esfahlani2021homogenization}. 
This coupling can be fully described by a polarizability tensor or scattering matrix representation, which relates the incident pressure and particle velocity to the induced responses and plays a pivotal role in governing the Willis lens. Under plane wave incidence and assuming an implicit time-harmonic convention $e^{-j\omega t}$, the linear acoustic wave equation can be written as follows \cite{wen2023acoustic,esfahlani2021homogenization}:
\begin{linenomath}
\begin{equation} \label{eq1}
\partial_z 
\begin{pmatrix}
p \\ v
\end{pmatrix}
=
j k
\begin{pmatrix}
\chi & \rho \\
\beta & \chi'
\end{pmatrix}
\begin{pmatrix}
p \\ v
\end{pmatrix}
\end{equation}
\end{linenomath}
where $k$ is the wavenumber, $p$ is the acoustic pressure, $v$ is the particle velocity, $\chi$ and $\chi'$ are the Willis coupling coefficients, and $z$ denotes the propagation direction (angle-dependent notation omitted for simplicity). We note that conventional symmetric media have zero Willis coupling coefficients. In addition, for reciprocal bianisotropic unit scatterers (e.g., lossless Willis particles), $\chi = \chi'$. On the other hand, non-Hermitian asymmetric particles, such as those employing viscous-thermal boundary effects in narrow passages (particularly in ultrasonic regimes), may have $\chi \neq \chi'$, indicating that the effective medium has nonreciprocal responses. Given the subwavelength nature of the proposed Willis scatterers, the scattering can be regarded as emanating from a small portion of the effective medium characterized by a lattice constant $d$. In this case, by integrating \cref{eq1} and utilizing the scattering (and/or transfer) matrix, we obtain the following Willis coupling coefficients:
\begin{linenomath}
\begin{equation} \label{eq2}
\begin{split}
\chi &= -\frac{2j}{\omega d}
   \frac{-1 + t_{11} - t_{12}\,t_{21} - t_{22} + t_{11}\,t_{22}}
              {1 + t_{11} - t_{12}\,t_{21} + t_{22} + t_{11}\,t_{22}},\\
\chi' &= -\frac{2j}{\omega d}
  \frac{-1 - t_{11} - t_{12}\,t_{21} + t_{22} + t_{11}\,t_{22}}
              {1 + t_{11} - t_{12}\,t_{21} + t_{22} + t_{11}\,t_{22}}, 
\end{split}
\end{equation}
\end{linenomath}
where $t_{kl}$ ($k,\,l=$ 1, 2) are the elements of the transfer matrix (see Supplementary Note 3 for the detailed derivation). We represent the Willis coupling coefficients extracted from the scattering responses of representative Willis scatterers, as shown in \cref{fig2}c, with their geometrical parameters indicated by star symbols in the bottom panel of \cref{fig2}d. The real (imaginary) parts of each coefficient are represented by solid (dashed) lines. It is well known that each unit cell exhibits the most dispersive characteristics and has maximum coupling coefficients near the local resonance frequency. Therefore, is has pronounced asymmetric reflection characteristics, whereas the transmission remains unaffected by the spatially asymmetric configuration (as shown in \cref{fig1}b). This allows for the realization of effective index modulation independent of the incident direction for the lens design. Moreover, the Willis coupling coefficients $\chi$ and $\chi'$ exhibit nearly identical values but opposite signs (\cref{fig2}c), that is, they have nearly reciprocal bianisotropic properties. This indicates that our system is nearly lossless owing to its low-frequency operation. By contrast, when thermoviscous losses become significant in the high-frequency regime, the system may exhibit nonreciprocal characteristics ($\chi \neq \chi'$). 

Without loss of generality, we design a spherical lens with a diameter of 240 mm by using the index profile of a generalized spherically symmetric lens, described by $n(r) = \sqrt{R^2 - r^2 +f^2 }/f$, where $R$ is the radius of the lens, and $f$ is the focal length. The design is based on a reference frequency of 31 kHz. We set the focal length to 0.2 m to prevent the focus from forming on the lens surface and for experimental verification. The desired effective index distribution and corresponding geometrical parameters are shown in \cref{fig2}d. Notably, Willis scatterers with resonance frequencies closer to the reference frequency are arranged nearer to the center of the lens. In addition, asymmetric backscattering is expected to be primarily governed by unit structures with strong pressure-velocity coupling, which are also located closer to the center of the lens.
\subsubsection*{Advantages of the Willis lens over conventional GRIN lenses}
Because Bragg scattering inevitably occurs at higher frequencies, conventional GRIN lenses inherently require a larger lattice constant than the Willis lens, which is locally resonant-based. This results in a lower spatial resolution or lower discretization levels, consequently reducing the focusing efficiency \cite{pan2022dielectric}. Moreover, although Willis scatterers can maintain consistent outer geometrical parameters ($r_\textnormal{c,o}$, $r_\textnormal{n,o}$), solid scatterers require larger geometrical parameters and consequently higher effective refractive indices as they approach the center of the lens, leading to an increased filling fraction approaching unity \cite{zhao2023review}. Furthermore, to ensure the overall structural stability of the lens, the size of the supporters connecting the unit cells must be increased, ultimately leading to an increase in the total mass.
\begin{figure}[!t]
\centering
\includegraphics [width=0.99\textwidth] {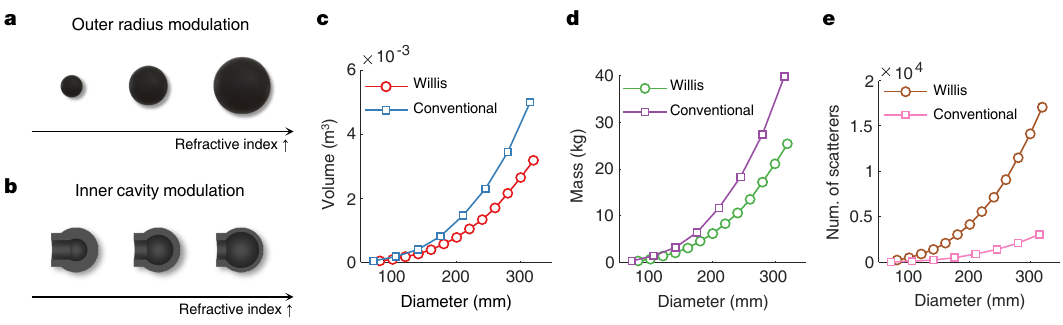}
\caption{\textbf{Advantages of Willis lenses over conventional underwater lenses.} 
\textbf{a, b} Schematic of refractive index modulation methods for (a) conventional and (b) Willis scatterers.
\textbf{c--e} Variation in the total volume (c), mass (d), and the number of unit scatterers (e) for the Willis and conventional lenses as a function of lens diameter. The Willis lens is significantly lighter, particularly at larger diameters, highlighting its weight reduction, cost, and scalability advantages.
}
\label{fig3}
\end{figure}
\Cref{fig3} depicts the changes in the volume and mass of a conventional (sphere-based) lens and a Willis lens as a function of the diameter designed for the same reference frequency. For this case, the lattice constant to design a conventional lens is set to 17.5 mm, which is 1.75 times that of our Willis scatterer. Interestingly, when considering lenses with a diameter of 280 mm, we find that despite having a larger number of unit structures\textemdash 11,513 for the Willis lens and 2,109 for the conventional one\textemdash the Willis lens is approximately 1.6 times lighter. 
In particular, the conventional lens weighs 27.5 kg. By contrast, the Willis lens weighs only 17.2 kg (\cref{fig3}d). This is because only the inner radius of the cavity is adjusted to achieve different refractive indices. By contrast, the outer geometrical parameters remain unchanged, resulting in a lightweight design while maintaining large discretization levels (\cref{fig3}a, b). 
Furthermore, the volume and mass differences between the conventional and Willis lenses becomes even more pronounced as the lens size increases. These findings highlight the superior advantages of Willis lenses, which achieve higher spatial resolution\textemdash leading to a higher focusing efficiency with reduced parasitic diffraction\textemdash and significantly reduce weight, making them more cost-effective. Therefore, they are a more appropriate choice for practical applications.
\subsubsection*{Experimental verification of underwater sound focusing with Willis lens}
\begin{figure}[!t]
\centering
\includegraphics [width=0.99\textwidth] {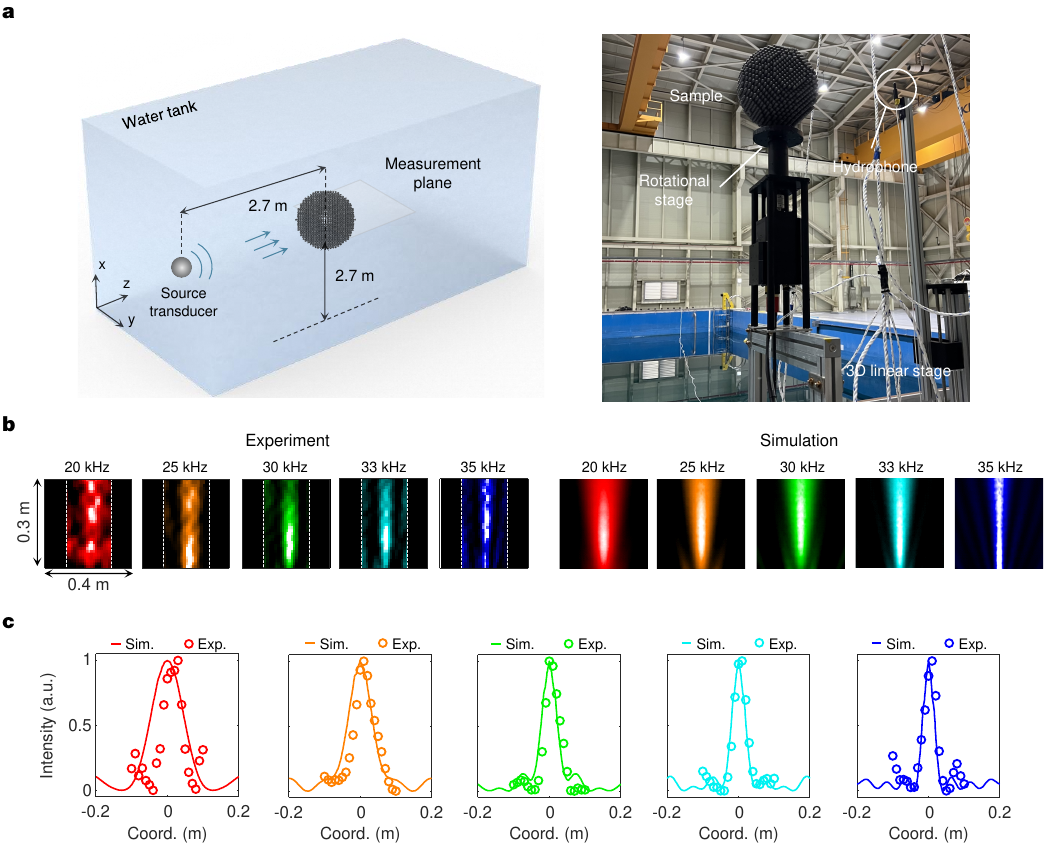}
\caption{\textbf{Experimental validation of broadband underwater focusing.} 
\textbf{a} Schematic (left) and photograph (right) of the experimental setup.
\textbf{b} Experimental (left panels) and numerical (right) intensity fields of the Willis lens in the $y$-$z$ plane at various frequencies indicated at the top. The incident wave propagates along the $z$-axis.
\textbf{c} One-dimensional point spread functions (PSFs) obtained from experimental (markers) and numerical (solid lines) results.
}
\label{fig4}
\end{figure}
A schematic of the experimental setup used to demonstrate the characteristics of the proposed Wills lens is shown in \cref{fig4}a. The measurements were conducted in a water tank (35 $\times$ 20 $\times$ 9.6 $\textnormal{m}^{3}$). The lens was fabricated as a monolithic structure by using additive manufacturing technology (see details in Methods and Supplementary Note 1). To the right of \cref{fig4}a, we show a photograph of the fabricated Willis lens mounted on the experimental setup (the input-source transducer is not indicated). The input waves are generated using an omnidirectional transducer, and the lens was positioned 2.7 m away from the source to ensure the incidence of a plane wavefront. A hydrophone scans the measurement plane (see \cref{fig4}a) near the focal point using point-by-point measurements with a three-dimensional automatic linear stage. A rotational stage was used for measurements at several incident directions. The dimension of the measurement area is 0.2 $\times$ 0.3 $\textnormal{m}^2$, and the pixel size is 0.01 $\times$ 0.01 $\textnormal{m}^2$ (approximately 0.18$\lambda$ $\times$ 0.18$\lambda$ based on the center frequency of 27.5 kHz). Further details of the experiments are provided in the Methods section. \Cref{fig4}b--c show the measured and calculated acoustic intensity distributions of the Willis lens at a $0^\circ$ angle of incidence (AOI). Overall, good agreement is observed between the captured and calculated results, demonstrating broadband focusing performance. The differences between the experimental and simulated results may be attributed to fabrication errors in the sample and undesired scattering caused by the experimental setup. Additional measurement results for the case without the lens are provided in Supplementary Note 5. 
Notably, the Willis lens maintains broadband focusing performance owing to the incidence-independent characteristics of our scatterers. This can be largely attributed to the three-dimensional nature of the Willis lens, which significantly outperforms conventional low-frequency underwater lenses \cite{su2017broadband,ruan20193}. Furthermore, the robust 3D focusing capability suggests that the Willis lens can achieve a higher SNR in submerged environments where ambient noise is dominant, making it a highly effective solution for practical applications. We also calculate the focusing gain of the sound pressure level (SPL) to quantify the performance of the Willis lens, obtaining an average SPL gain of 8.07 dB and 11.09 dB for the experimental and numerical results, respectively, indicating robust focusing performance (see details in Supplementary Note 6).
\subsubsection*{Wide-angle focusing with asymmetric scattering characteristics}
\begin{figure}[!h]
\centering
\includegraphics [width=0.99\textwidth] {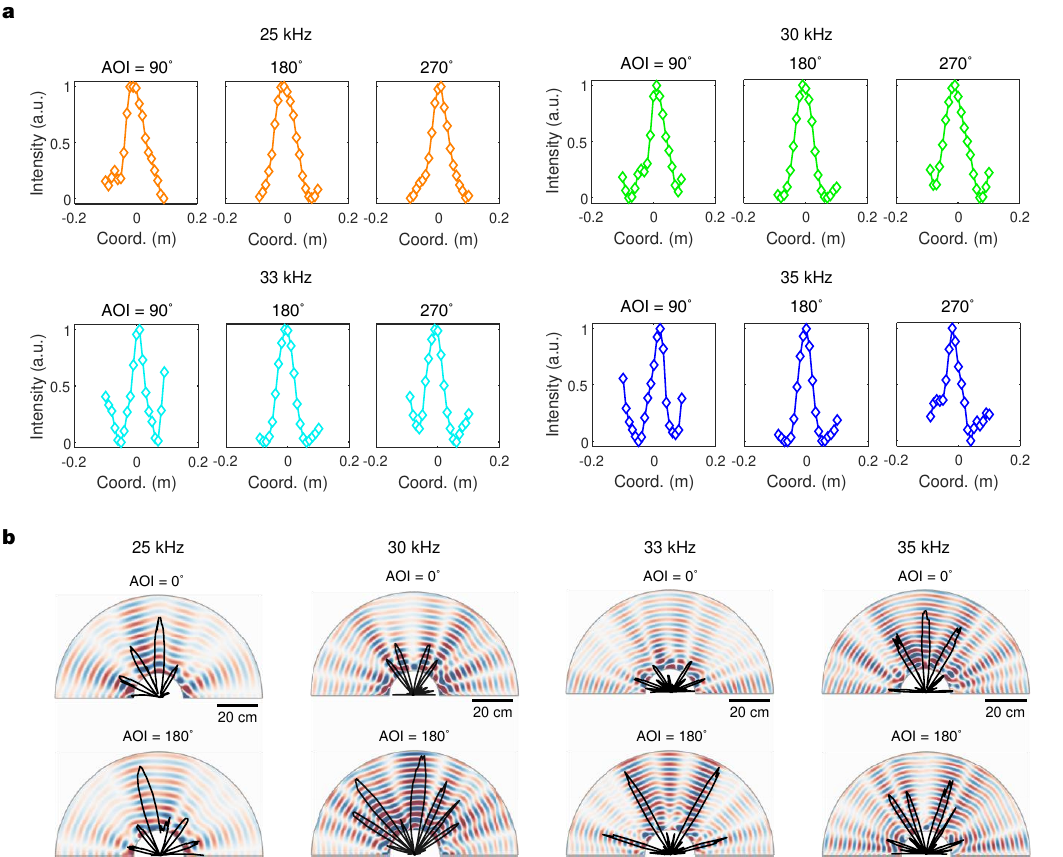}
\caption{\textbf{Incidence-dependent and -independent characteristics of the Willis lens.}
\textbf{a} Measured intensity profiles for AOIs of 90°, 180°, and 270°, and frequencies of 25, 30, 33, and 35 kHz.
\textbf{b} Representation of asymmetric backscattering fields and scattering patterns for different incidence directions. In the case of a $0^{\circ}$ AOI, the neck region of the Willis scatterer points toward the propagation direction, while at 180°, the direction is reversed. The scale bar represents a length of 20 cm.
}
\label{fig5}
\end{figure}
We now investigate the details of incidence-dependent asymmetric backscattering and incidence-independent wide-angle focusing characteristics of the Willis lens. \Cref{fig5}a shows the measured one-dimensional point spread functions (PSFs) at various frequencies. As aforedescribed, our Willis lens exhibits incident-independent transmission and incident-dependent reflection responses. Therefore, we can observe that regardless of the incident directions, the Willis lens maintains consistent focusing field distributions, making it suitable for wide-field underwater focusing (see Supplementary Note 5 for additional experimental results). \Cref{fig5}b shows the scattering patterns and field distributions for different lens configurations. To enhance visualization, we overlay the semicircular beam pattern and field distribution, representing the region opposite to the focusing direction. As previously discussed, Willis scatterers exhibit asymmetric reflection phases across the entire operational frequency range, with each scatterer reaching maximum Willis coupling near its locally resonant frequency. Consequently, asymmetric backscattering responses were observed across a wide frequency range. Interestingly, for mirror-symmetric configurations, such as $90^{\circ}$ and $270^{\circ}$ AOIs, we observe that the scattering fields are nearly symmetric (see Supplementary Note 7 for details). Based on these characteristics, we anticipate that increasing the design degree of freedom of Willis scatterers would enable the simultaneous control of the focusing and backscattering properties.
\section*{Discussion}
We devised a new design for underwater acoustic focusing, diverging from traditional solid scatterer-based design approaches, and demonstrated an efficient solution for practical applications. 
Our results open new possibilities for expanding into the lower-frequency range, including the audible spectrum, which becomes increasingly significant in submerged environments. The experimental results demonstrate the robustness of the focusing performance in real-world applications. 
Notably, the proposed Willis lens enables significant weight reduction while maintaining a robust focusing performance, showcasing its potential for practical applications. 
Additionally, we investigated the asymmetric characteristics of the Willis lens, which arise owing to its bianisotropic nature, resulting from the spatially asymmetric configuration. While this work primarily focused on broadband focusing performance rather than a detailed design approach for artificially controlling backscattering fields, further development is anticipated, such as the realization of dual functionality\textemdash simultaneous focusing and diffusion-like backscattering using our Willis scatterers. This study essentially serves as a starting point for further investigations that will lead to the development of robust platforms for on-demand waterborne sound focusing and their integration into underwater wireless sensor networks.
Future research should address the acoustic (dynamic) properties and structural (static) characteristics, including varying hydrostatic pressures at different submerged depths and drag forces from ocean currents during vessel maneuvers. This integrated consideration of dynamic and static factors will establish a substantial foundation for underwater lens design, marking a promising direction for future studies.

\section*{Methods}
\subsubsection*{Numerical simulations}
The full-wave simulations were conducted using a finite-element method based on the commercial software, \textsc{Comsol Multiphysics 6.2}. The mass density and sound speed of the background medium (water) are 999 $\textnormal{kg} \cdot \textnormal{m}^{-3}$ and 1460 m$\cdot \textnormal{s}^{-1}$, respectively. The lens structure was assigned a mass density of 7980 $\textnormal{kg} \cdot \textnormal{m}^{-3}$, a Young's modulus of 193 $ \textnormal{GPa} $ and a Poisson ratio of 0.3. 
A half-model of the proposed Willis lens was analyzed using a three-dimensional simulation by applying symmetry conditions to reduce computational cost. To simulate realistic conditions, a background pressure field was implemented, accompanied by a plane wave radiation condition at the output boundaries. Additionally, cylindrical wave radiation conditions were applied along the lateral boundaries of the domain to accurately represent free-field wave propagation, minimizing unwanted reflection and diffraction.

\subsubsection*{Sample fabrication}
Our Willis lens was fabricated using a metal additive printer (Concept laser M2 cusing; GE Additive). The base powder used for lens fabrication is STS 316L, which offers resistance to rust and degradation, even in saline or harsh underwater environments. Lens supporters are affixed to the underside of the lens to ensure proper integration with the measurement system. After completing the structure printing, various post-processing steps were performed, including laser cutting, hot isostatic pressing (HIP), and other treatments designed to improve the overall quality and structural reliability of the fabricated sample. Finally, surface treatment processes were carried out to ensure uniform surface roughness across the sample, ensuring its mechanical properties.
To validate whether the fabricated sample was manufactured with the desired properties e.g., Young's modulus, we conducted a vibration test using the standard ASTM E1876 (resonance method, see Supplementary Note 2 for details).

\subsubsection*{Experiments}
All experiments were conducted in a water tank with dimensions of 35 (length) × 20 (width) × 9.6 (depth) $\textnormal{m}^{3}$, located at Korea Institute of Ocean Science \& Technology (KIOST). The fabricated sample was submerged by a hoist system. A omnidirectional projector (ITC-1042; frequency range 0.01 to 100 kHz; International Transducer Corporation) was placed at 2.7 m in front of the lens to generate incidence plane waves (see details in Supplementary Note 5). The input signal was applied to the projector using a arbitrary waveform generator (PXI-5421; National Instruments) and power amplifier (Type 2713; Brüel \& Kjær). For signal transmission, a series of burst tones within the 20–35 kHz frequency range was employed. To perform sound field scanning at the measuring planes, a hydrophone (Type 8103; Brüel \& Kjær) was employed. After calibration with a hydrophone calibrator (Type 4229; Brüel \& Kjær), the hydrophone was mounted on an automated 3D positioning system and used as the receiver to detect and record acoustic pressure with high precision. The scanning resolution was set to 1 cm, with a 2 s interval between each measurement. The output acoustic signals were conditioned by a charge amplifier (Type 2692; Brüel \& Kjær). We measured the sound velocity and temperature using the sound velocity profiler (SWiFT SVP; Valeport) in the experimental environment. This was used to calculate the arrival times from the source to the lens (see details in Supplementary Note 4).

\backmatter

\nolinenumbers

\bibliography{Reference}

\begin{thebibliography}{10}
\expandafter\ifx\csname url\endcsname\relax
  \def\url#1{\burl{#1}}\fi
\expandafter\ifx\csname urlprefix\endcsname\relax\def\urlprefix{URL }\fi
\providecommand{\bibinfo}[2]{#2}
\providecommand{\eprint}[2][]{\url{#2}}
\providecommand{\doi}[1]{\url{https://doi.org/#1}}
\bibcommenthead

\bibitem{akyildiz2005underwater}
\bibinfo{author}{Akyildiz, I.~F.}, \bibinfo{author}{Pompili, D.} \& \bibinfo{author}{Melodia, T.}
\newblock \bibinfo{title}{Underwater acoustic sensor networks: research challenges}.
\newblock \emph{\bibinfo{journal}{Ad hoc networks}} \textbf{\bibinfo{volume}{3}}, \bibinfo{pages}{257--279} (\bibinfo{year}{2005}).

\bibitem{felemban2015underwater}
\bibinfo{author}{Felemban, E.}, \bibinfo{author}{Shaikh, F.~K.}, \bibinfo{author}{Qureshi, U.~M.}, \bibinfo{author}{Sheikh, A.~A.} \& \bibinfo{author}{Qaisar, S.~B.}
\newblock \bibinfo{title}{Underwater sensor network applications: A comprehensive survey}.
\newblock \emph{\bibinfo{journal}{International Journal of Distributed Sensor Networks}} \textbf{\bibinfo{volume}{11}}, \bibinfo{pages}{896832} (\bibinfo{year}{2015}).

\bibitem{elfes1987sonar}
\bibinfo{author}{Elfes, A.}
\newblock \bibinfo{title}{Sonar-based real-world mapping and navigation}.
\newblock \emph{\bibinfo{journal}{IEEE Journal on Robotics and Automation}} \textbf{\bibinfo{volume}{3}}, \bibinfo{pages}{249--265} (\bibinfo{year}{1987}).

\bibitem{andre2011listening}
\bibinfo{author}{Andr{\'e}, M.} \emph{et~al.}
\newblock \bibinfo{title}{Listening to the deep: live monitoring of ocean noise and cetacean acoustic signals}.
\newblock \emph{\bibinfo{journal}{Marine pollution bulletin}} \textbf{\bibinfo{volume}{63}}, \bibinfo{pages}{18--26} (\bibinfo{year}{2011}).

\bibitem{hovem2007underwater}
\bibinfo{author}{Hovem, J.~M.}
\newblock \bibinfo{title}{Underwater acoustics: Propagation, devices and systems}.
\newblock \emph{\bibinfo{journal}{Journal of Electroceramics}} \textbf{\bibinfo{volume}{19}}, \bibinfo{pages}{339--347} (\bibinfo{year}{2007}).

\bibitem{sherman2007transducers}
\bibinfo{author}{Sherman, C.~H.} \& \bibinfo{author}{Butler, J.~L.}
\newblock \emph{\bibinfo{title}{Transducers and arrays for underwater sound}} Vol.~\bibinfo{volume}{4} (\bibinfo{publisher}{Springer}, \bibinfo{year}{2007}).

\bibitem{afzal2022battery}
\bibinfo{author}{Afzal, S.~S.} \emph{et~al.}
\newblock \bibinfo{title}{Battery-free wireless imaging of underwater environments}.
\newblock \emph{\bibinfo{journal}{Nature communications}} \textbf{\bibinfo{volume}{13}}, \bibinfo{pages}{5546} (\bibinfo{year}{2022}).

\bibitem{cummer2016controlling}
\bibinfo{author}{Cummer, S.~A.}, \bibinfo{author}{Christensen, J.} \& \bibinfo{author}{Al{\`u}, A.}
\newblock \bibinfo{title}{Controlling sound with acoustic metamaterials}.
\newblock \emph{\bibinfo{journal}{Nature Reviews Materials}} \textbf{\bibinfo{volume}{1}}, \bibinfo{pages}{1--13} (\bibinfo{year}{2016}).

\bibitem{assouar2018acoustic}
\bibinfo{author}{Assouar, B.} \emph{et~al.}
\newblock \bibinfo{title}{Acoustic metasurfaces}.
\newblock \emph{\bibinfo{journal}{Nature Reviews Materials}} \textbf{\bibinfo{volume}{3}}, \bibinfo{pages}{460--472} (\bibinfo{year}{2018}).

\bibitem{oh2023engineering}
\bibinfo{author}{Oh, B.}, \bibinfo{author}{Kim, K.}, \bibinfo{author}{Lee, D.} \& \bibinfo{author}{Rho, J.}
\newblock \bibinfo{title}{Engineering metalenses for planar optics and acoustics}.
\newblock \emph{\bibinfo{journal}{Materials Today Physics}} \bibinfo{pages}{101273} (\bibinfo{year}{2023}).

\bibitem{lee2022piezoelectric}
\bibinfo{author}{Lee, G.} \emph{et~al.}
\newblock \bibinfo{title}{Piezoelectric energy harvesting using mechanical metamaterials and phononic crystals}.
\newblock \emph{\bibinfo{journal}{Communications Physics}} \textbf{\bibinfo{volume}{5}}, \bibinfo{pages}{94} (\bibinfo{year}{2022}).

\bibitem{vasileiadis2021progress}
\bibinfo{author}{Vasileiadis, T.} \emph{et~al.}
\newblock \bibinfo{title}{Progress and perspectives on phononic crystals}.
\newblock \emph{\bibinfo{journal}{Journal of Applied Physics}} \textbf{\bibinfo{volume}{129}} (\bibinfo{year}{2021}).

\bibitem{lee2024wide}
\bibinfo{author}{Lee, D.} \emph{et~al.}
\newblock \bibinfo{title}{Wide field-of-hearing metalens for aberration-free sound capture}.
\newblock \emph{\bibinfo{journal}{Nature Communications}} \textbf{\bibinfo{volume}{15}}, \bibinfo{pages}{3044} (\bibinfo{year}{2024}).

\bibitem{dong2023underwater}
\bibinfo{author}{Dong, E.} \emph{et~al.}
\newblock \bibinfo{title}{Underwater acoustic metamaterials}.
\newblock \emph{\bibinfo{journal}{National Science Review}} \textbf{\bibinfo{volume}{10}}, \bibinfo{pages}{nwac246} (\bibinfo{year}{2023}).

\bibitem{zhou2024underwater}
\bibinfo{author}{Zhou, H.-T.} \emph{et~al.}
\newblock \bibinfo{title}{Underwater scattering exceptional point by metasurface with fluid-solid interaction}.
\newblock \emph{\bibinfo{journal}{Advanced Functional Materials}} \bibinfo{pages}{2404282} (\bibinfo{year}{2024}).

\bibitem{li2024janus}
\bibinfo{author}{Li, C.-Y.}, \bibinfo{author}{Zhou, H.-T.}, \bibinfo{author}{Li, X.-S.}, \bibinfo{author}{Wang, Y.-F.} \& \bibinfo{author}{Wang, Y.-S.}
\newblock \bibinfo{title}{Janus metasurface for underwater sound manipulation}.
\newblock \emph{\bibinfo{journal}{Advanced Functional Materials}} \textbf{\bibinfo{volume}{34}}, \bibinfo{pages}{2408572} (\bibinfo{year}{2024}).

\bibitem{qu2022underwater}
\bibinfo{author}{Qu, S.} \emph{et~al.}
\newblock \bibinfo{title}{Underwater metamaterial absorber with impedance-matched composite}.
\newblock \emph{\bibinfo{journal}{Science Advances}} \textbf{\bibinfo{volume}{8}}, \bibinfo{pages}{eabm4206} (\bibinfo{year}{2022}).

\bibitem{lee2021underwater}
\bibinfo{author}{Lee, D.} \emph{et~al.}
\newblock \bibinfo{title}{Underwater stealth metasurfaces composed of split-orifice--conduit hybrid resonators}.
\newblock \emph{\bibinfo{journal}{Journal of Applied Physics}} \textbf{\bibinfo{volume}{129}} (\bibinfo{year}{2021}).

\bibitem{yang2023experimental}
\bibinfo{author}{Yang, T.}, \bibinfo{author}{Lin, Z.} \& \bibinfo{author}{Yang, T.}
\newblock \bibinfo{title}{Experimental evidence of high-efficiency nonlocal waterborne acoustic metasurfaces}.
\newblock \emph{\bibinfo{journal}{Advanced Engineering Materials}} \textbf{\bibinfo{volume}{25}}, \bibinfo{pages}{2200805} (\bibinfo{year}{2023}).

\bibitem{zhao2023review}
\bibinfo{author}{Zhao, L.}, \bibinfo{author}{Bi, C.}, \bibinfo{author}{Huang, H.}, \bibinfo{author}{Liu, Q.} \& \bibinfo{author}{Tian, Z.}
\newblock \bibinfo{title}{A review of acoustic luneburg lens: Physics and applications}.
\newblock \emph{\bibinfo{journal}{Mechanical Systems and Signal Processing}} \textbf{\bibinfo{volume}{199}}, \bibinfo{pages}{110468} (\bibinfo{year}{2023}).

\bibitem{luneburg1966mathematical}
\bibinfo{author}{Luneburg, R.~K.}
\newblock \emph{\bibinfo{title}{Mathematical theory of optics}}  (\bibinfo{publisher}{Univ of California Press}, \bibinfo{year}{1966}).

\bibitem{allam20203d}
\bibinfo{author}{Allam, A.}, \bibinfo{author}{Sabra, K.} \& \bibinfo{author}{Erturk, A.}
\newblock \bibinfo{title}{3d-printed gradient-index phononic crystal lens for underwater acoustic wave focusing}.
\newblock \emph{\bibinfo{journal}{Physical Review Applied}} \textbf{\bibinfo{volume}{13}}, \bibinfo{pages}{064064} (\bibinfo{year}{2020}).

\bibitem{kim2022three}
\bibinfo{author}{Kim, J.-W.}, \bibinfo{author}{Hwang, G.}, \bibinfo{author}{Lee, S.-J.}, \bibinfo{author}{Kim, S.-H.} \& \bibinfo{author}{Wang, S.}
\newblock \bibinfo{title}{Three-dimensional acoustic metamaterial luneburg lenses for broadband and wide-angle underwater ultrasound imaging}.
\newblock \emph{\bibinfo{journal}{Mechanical Systems and Signal Processing}} \textbf{\bibinfo{volume}{179}}, \bibinfo{pages}{109374} (\bibinfo{year}{2022}).

\bibitem{xie2018acoustic}
\bibinfo{author}{Xie, Y.} \emph{et~al.}
\newblock \bibinfo{title}{Acoustic imaging with metamaterial luneburg lenses}.
\newblock \emph{\bibinfo{journal}{Scientific reports}} \textbf{\bibinfo{volume}{8}}, \bibinfo{pages}{16188} (\bibinfo{year}{2018}).

\bibitem{li2021focus}
\bibinfo{author}{Li, Z.} \emph{et~al.}
\newblock \bibinfo{title}{Focus of ultrasonic underwater sound with 3d printed phononic crystal}.
\newblock \emph{\bibinfo{journal}{Applied Physics Letters}} \textbf{\bibinfo{volume}{119}} (\bibinfo{year}{2021}).

\bibitem{lu2021grin}
\bibinfo{author}{Lu, C.} \emph{et~al.}
\newblock \bibinfo{title}{Grin metamaterial generalized luneburg lens for ultra-long acoustic jet}.
\newblock \emph{\bibinfo{journal}{Applied Physics Letters}} \textbf{\bibinfo{volume}{118}} (\bibinfo{year}{2021}).

\bibitem{kim2021acoustic}
\bibinfo{author}{Kim, J.-W.}, \bibinfo{author}{Lee, S.-J.}, \bibinfo{author}{Jo, J.-Y.}, \bibinfo{author}{Wang, S.} \& \bibinfo{author}{Kim, S.-H.}
\newblock \bibinfo{title}{Acoustic imaging by three-dimensional acoustic luneburg meta-lens with lattice columns}.
\newblock \emph{\bibinfo{journal}{Applied Physics Letters}} \textbf{\bibinfo{volume}{118}} (\bibinfo{year}{2021}).

\bibitem{kim2021poroelastic}
\bibinfo{author}{Kim, G.}, \bibinfo{author}{Portela, C.~M.}, \bibinfo{author}{Celli, P.}, \bibinfo{author}{Palermo, A.} \& \bibinfo{author}{Daraio, C.}
\newblock \bibinfo{title}{Poroelastic microlattices for underwater wave focusing}.
\newblock \emph{\bibinfo{journal}{Extreme Mechanics Letters}} \textbf{\bibinfo{volume}{49}}, \bibinfo{pages}{101499} (\bibinfo{year}{2021}).

\bibitem{tong20233d}
\bibinfo{author}{Tong, S.} \& \bibinfo{author}{Ren, C.}
\newblock \bibinfo{title}{3d underwater acoustic luneburg lens based on gradient face-centered-cubic phononic crystals}.
\newblock \emph{\bibinfo{journal}{Applied Physics Letters}} \textbf{\bibinfo{volume}{123}} (\bibinfo{year}{2023}).

\bibitem{quan2018maximum}
\bibinfo{author}{Quan, L.}, \bibinfo{author}{Ra’di, Y.}, \bibinfo{author}{Sounas, D.~L.} \& \bibinfo{author}{Al{\`u}, A.}
\newblock \bibinfo{title}{Maximum willis coupling in acoustic scatterers}.
\newblock \emph{\bibinfo{journal}{Physical review letters}} \textbf{\bibinfo{volume}{120}}, \bibinfo{pages}{254301} (\bibinfo{year}{2018}).

\bibitem{sieck2017origins}
\bibinfo{author}{Sieck, C.~F.}, \bibinfo{author}{Al{\`u}, A.} \& \bibinfo{author}{Haberman, M.~R.}
\newblock \bibinfo{title}{Origins of willis coupling and acoustic bianisotropy in acoustic metamaterials through source-driven homogenization}.
\newblock \emph{\bibinfo{journal}{Physical Review B}} \textbf{\bibinfo{volume}{96}}, \bibinfo{pages}{104303} (\bibinfo{year}{2017}).

\bibitem{esfahlani2021homogenization}
\bibinfo{author}{Esfahlani, H.}, \bibinfo{author}{Mazor, Y.} \& \bibinfo{author}{Al{\`u}, A.}
\newblock \bibinfo{title}{Homogenization and design of acoustic willis metasurfaces}.
\newblock \emph{\bibinfo{journal}{Physical Review B}} \textbf{\bibinfo{volume}{103}}, \bibinfo{pages}{054306} (\bibinfo{year}{2021}).

\bibitem{muhlestein2017experimental}
\bibinfo{author}{Muhlestein, M.~B.}, \bibinfo{author}{Sieck, C.~F.}, \bibinfo{author}{Wilson, P.~S.} \& \bibinfo{author}{Haberman, M.~R.}
\newblock \bibinfo{title}{Experimental evidence of willis coupling in a one-dimensional effective material element}.
\newblock \emph{\bibinfo{journal}{Nature communications}} \textbf{\bibinfo{volume}{8}}, \bibinfo{pages}{15625} (\bibinfo{year}{2017}).

\bibitem{melnikov2019acoustic}
\bibinfo{author}{Melnikov, A.} \emph{et~al.}
\newblock \bibinfo{title}{Acoustic meta-atom with experimentally verified maximum willis coupling}.
\newblock \emph{\bibinfo{journal}{Nature communications}} \textbf{\bibinfo{volume}{10}}, \bibinfo{pages}{3148} (\bibinfo{year}{2019}).

\bibitem{su2018retrieval}
\bibinfo{author}{Su, X.} \& \bibinfo{author}{Norris, A.~N.}
\newblock \bibinfo{title}{Retrieval method for the bianisotropic polarizability tensor of willis acoustic scatterers}.
\newblock \emph{\bibinfo{journal}{Physical Review B}} \textbf{\bibinfo{volume}{98}}, \bibinfo{pages}{174305} (\bibinfo{year}{2018}).

\bibitem{sepehrirahnama2022willis}
\bibinfo{author}{Sepehrirahnama, S.}, \bibinfo{author}{Oberst, S.}, \bibinfo{author}{Chiang, Y.~K.} \& \bibinfo{author}{Powell, D.~A.}
\newblock \bibinfo{title}{Willis coupling-induced acoustic radiation force and torque reversal}.
\newblock \emph{\bibinfo{journal}{Physical Review Letters}} \textbf{\bibinfo{volume}{129}}, \bibinfo{pages}{174501} (\bibinfo{year}{2022}).

\bibitem{li2018systematic}
\bibinfo{author}{Li, J.}, \bibinfo{author}{Shen, C.}, \bibinfo{author}{D{\'\i}az-Rubio, A.}, \bibinfo{author}{Tretyakov, S.~A.} \& \bibinfo{author}{Cummer, S.~A.}
\newblock \bibinfo{title}{Systematic design and experimental demonstration of bianisotropic metasurfaces for scattering-free manipulation of acoustic wavefronts}.
\newblock \emph{\bibinfo{journal}{Nature communications}} \textbf{\bibinfo{volume}{9}}, \bibinfo{pages}{1342} (\bibinfo{year}{2018}).

\bibitem{muhlestein2016reciprocity}
\bibinfo{author}{Muhlestein, M.~B.}, \bibinfo{author}{Sieck, C.~F.}, \bibinfo{author}{Al{\`u}, A.} \& \bibinfo{author}{Haberman, M.~R.}
\newblock \bibinfo{title}{Reciprocity, passivity and causality in willis materials}.
\newblock \emph{\bibinfo{journal}{Proceedings of the Royal Society A: Mathematical, Physical and Engineering Sciences}} \textbf{\bibinfo{volume}{472}}, \bibinfo{pages}{20160604} (\bibinfo{year}{2016}).

\bibitem{demir2024equivalence}
\bibinfo{author}{Demir, M.~U.} \& \bibinfo{author}{Popa, B.-I.}
\newblock \bibinfo{title}{Equivalence between general acoustic willis media and conventional materials with embedded sources}.
\newblock \emph{\bibinfo{journal}{Physical Review B}} \textbf{\bibinfo{volume}{109}}, \bibinfo{pages}{L020301} (\bibinfo{year}{2024}).

\bibitem{wen2023acoustic}
\bibinfo{author}{Wen, X.}, \bibinfo{author}{Yip, H.~K.}, \bibinfo{author}{Cho, C.}, \bibinfo{author}{Li, J.} \& \bibinfo{author}{Park, N.}
\newblock \bibinfo{title}{Acoustic amplifying diode using nonreciprocal willis coupling}.
\newblock \emph{\bibinfo{journal}{Physical Review Letters}} \textbf{\bibinfo{volume}{130}}, \bibinfo{pages}{176101} (\bibinfo{year}{2023}).

\bibitem{asadchy2018bianisotropic}
\bibinfo{author}{Asadchy, V.~S.}, \bibinfo{author}{D{\'\i}az-Rubio, A.} \& \bibinfo{author}{Tretyakov, S.~A.}
\newblock \bibinfo{title}{Bianisotropic metasurfaces: physics and applications}.
\newblock \emph{\bibinfo{journal}{Nanophotonics}} \textbf{\bibinfo{volume}{7}}, \bibinfo{pages}{1069--1094} (\bibinfo{year}{2018}).

\bibitem{kriegler2009bianisotropic}
\bibinfo{author}{Kriegler, C.~E.}, \bibinfo{author}{Rill, M.~S.}, \bibinfo{author}{Linden, S.} \& \bibinfo{author}{Wegener, M.}
\newblock \bibinfo{title}{Bianisotropic photonic metamaterials}.
\newblock \emph{\bibinfo{journal}{IEEE journal of selected topics in quantum electronics}} \textbf{\bibinfo{volume}{16}}, \bibinfo{pages}{367--375} (\bibinfo{year}{2009}).

\bibitem{pan2022dielectric}
\bibinfo{author}{Pan, M.} \emph{et~al.}
\newblock \bibinfo{title}{Dielectric metalens for miniaturized imaging systems: progress and challenges}.
\newblock \emph{\bibinfo{journal}{Light: Science \& Applications}} \textbf{\bibinfo{volume}{11}}, \bibinfo{pages}{195} (\bibinfo{year}{2022}).

\bibitem{liu2000locally}
\bibinfo{author}{Liu, Z.} \emph{et~al.}
\newblock \bibinfo{title}{Locally resonant sonic materials}.
\newblock \emph{\bibinfo{journal}{science}} \textbf{\bibinfo{volume}{289}}, \bibinfo{pages}{1734--1736} (\bibinfo{year}{2000}).

\bibitem{ma2016acoustic}
\bibinfo{author}{Ma, G.} \& \bibinfo{author}{Sheng, P.}
\newblock \bibinfo{title}{Acoustic metamaterials: From local resonances to broad horizons}.
\newblock \emph{\bibinfo{journal}{Science advances}} \textbf{\bibinfo{volume}{2}}, \bibinfo{pages}{e1501595} (\bibinfo{year}{2016}).

\bibitem{su2017broadband}
\bibinfo{author}{Su, X.}, \bibinfo{author}{Norris, A.~N.}, \bibinfo{author}{Cushing, C.~W.}, \bibinfo{author}{Haberman, M.~R.} \& \bibinfo{author}{Wilson, P.~S.}
\newblock \bibinfo{title}{Broadband focusing of underwater sound using a transparent pentamode lens}.
\newblock \emph{\bibinfo{journal}{The Journal of the Acoustical Society of America}} \textbf{\bibinfo{volume}{141}}, \bibinfo{pages}{4408--4417} (\bibinfo{year}{2017}).

\bibitem{ruan20193}
\bibinfo{author}{Ruan, Y.} \emph{et~al.}
\newblock \bibinfo{title}{3-d underwater acoustic wave focusing by periodic structure}.
\newblock \emph{\bibinfo{journal}{Applied Physics Letters}} \textbf{\bibinfo{volume}{114}} (\bibinfo{year}{2019}).

\end{thebibliography}
\section*{Data availability}
Data that support the findings of this study are available from the corresponding authors upon request.

\section*{Acknowledgments}
This work was financially supported by the POSCO-POSTECH-RIST Convergence Research Center program funded by POSCO, the National Research Foundation (NRF) grants (RS-2024-00356928, RS-2024-00436476) funded by the Ministry of Science and ICT (MSIT) of the Korean government, and the grant (PES5550) from the endowment project of “Development of smart sensor technology for underwater environment monitoring” funded by Korea Research Institute of Ships \& Ocean engineering (KRISO). B.O. acknowledges the NRF Ph.D. fellowship (RS-2024-00409956) funded by the Ministry of Education of the Korean government. The authors acknowledge N.-Y. Yun, J.B. Jang and J. Choi for their help with acoustic field and sound velocity profile measurements.

\section*{Author contributions}
B.O. and J.R. conceived the idea and initiated the project. B.O. and D.L. performed theoretical analyses. B.O. performed numerical simulations. B.O., S.-M.K. and Y.-S.C. performed the experiments. S.-H.B. and J.S. supported the experiments. S.-M.K. supported the data analyses. All authors participated in discussions and confirmed the final manuscript. J.R. and S.-M.K. guided the entire work.

\section*{Competing interests} The authors declare no competing interests.

\end{document}